\documentclass[reprint,noshowpacs,noshowkeys,prd,balancelastpage,nofootinbib]{revtex4}
\usepackage{amsfonts}
\usepackage{xcolor}
\usepackage{mdframed}
\usepackage{amssymb}
\usepackage{footnote}
\usepackage{amsmath}
\usepackage{graphicx}
\usepackage{float}
\usepackage[font={footnotesize,it}]{caption}
\usepackage[utf8]{inputenc}
\usepackage{natbib}
\usepackage{tikz}
\usepackage{tikz-3dplot}
\usepackage[colorlinks=true,
            linkcolor=purple,
            urlcolor=purple,
            citecolor=blue]{hyperref}

\begin{document}

\title{Charged quantum Oppenheimer-Snyder model}
\author{S. Habib Mazharimousavi}
\email{habib.mazhari@emu.edu.tr}
\affiliation{Department of Physics, Faculty of Arts and Sciences, Eastern Mediterranean
University, Famagusta, North Cyprus via Mersin 10, T\"{u}rkiye}
\date{\today }

\begin{abstract}
In the framework of loop quantum cosmology, particularly within the quantum
Oppenheimer-Snyder model, the semiclassical Ashtekar-Pawlowski-Singh (APS)
metric is associated with a static, spherically symmetric black hole that
incorporates quantum effects derived from the APS metric. This
quantum-corrected black hole can be interpreted as a modified Schwarzschild
black hole, where the Schwarzschild metric function is adjusted by an
additional term proportional to $\frac{M^{2}}{r^{4}}$, with $r$ denoting the
radial coordinate and $M,$ the black hole mass. In this study, we show that
such a quantum-mechanically modified black hole can arise in the context of
nonlinear electrodynamics with either electric or magnetic charge. This
charged, quantum-corrected solution is then matched to a dust ball of
constant mass $M_{APS}$, governed by the APS metric, at a timelike
thin-shell possessing nonzero mass $m$ and electric charge $Q$ or magnetic
charge $P$. Analytically, it is demonstrated that the thin-shell oscillates
around an equilibrium radius $r=R_{eq}$, which is expressed in terms of $%
M_{APS}$, $m$, and $Q$ or $P$.
\end{abstract}

\maketitle

\section{Introduction}

Recently, in \cite{1}, a nonsingular black hole was proposed within the
framework of loop quantum gravity that is called the quantum
Oppenheimer-Snyder (qOS) model. This model of nonsingular black hole is
composed of two distinct regions - the interior and the exterior - which are
joined together at a timelike spherical interface hypersurface. The interior
region is a dust ball described by the semiclassical
Ashtekar-Pawlowski-Singh (APS) metric \cite{APS}, with the line element 
\begin{equation}
ds_{APS}^{2}=-d\tau ^{2}+a\left( \tau \right) ^{2}\left\{ d\tilde{r}^{2}+%
\tilde{r}^{2}\left( d\theta ^{2}+\sin ^{2}\theta d\phi ^{2}\right) \right\} ,
\label{1}
\end{equation}%
which applies to the region inside the hypersurface $\tilde{r}=\tilde{r}_{0}$%
. Here, the expansion parameter $a\left( \tau \right) $ satisfies the
modified Friedmann equation given by

\begin{equation}
\left( \frac{\dot{a}}{a}\right) ^{2}=\frac{8\pi G}{3}\rho \left( 1-\frac{%
\rho }{\rho _{c}}\right) ,  \label{2}
\end{equation}%
where $\rho $ is the energy density and $\rho _{c}$ is the critical energy
density. Additionally, the conservation of energy, $\nabla _{\mu }T^{\mu \nu
}=0$ with 
\begin{equation}
T^{\mu \nu }=diag\left[ \rho ,0,0,0\right] ,  \label{K1}
\end{equation}%
implies 
\begin{equation}
\frac{d\rho }{d\tau }+\frac{3\dot{a}}{a}\rho =0,  \label{3}
\end{equation}%
leading to%
\begin{equation}
\rho =\frac{M}{\frac{4}{3}\pi \tilde{r}_{0}^{3}a^{3}},  \label{4}
\end{equation}%
where $M$ represents the total constant mass of the interior spacetime,
which extends from $r=0$ to $r=\tilde{r}_{0}a$. In (\ref{2}), $\rho _{c}$ is
the critical energy density such that in the classical regime (i.e., $\rho
\leq \rho _{c}$), the modified Friedmann equation (\ref{2}) reduces to the
standard Friedmann equation. Conversely, in the quantum regime where $\rho $
becomes comparable to $\rho _{c}$, the term $\frac{\rho }{\rho _{c}}$ in (%
\ref{2}) prevents the energy density from diverging. This can be verified by
substituting Eq. (\ref{4}) into Eq. (\ref{2}) and solving for $a$. Although
Eq. (\ref{2}) is solvable, its structure indicates that $\rho _{c}$
represents the maximum attainable energy density for $\rho $.

The exterior spacetime is described by a quantum-mechanically corrected
black hole, with the line element 
\begin{equation}
ds^{2}=-\left( 1-\frac{2M}{r}+\frac{\alpha M^{2}}{r^{4}}\right) dt^{2}+\frac{%
dr^{2}}{1-\frac{2M}{r}+\frac{\alpha M^{2}}{r^{4}}}+r^{2}\left( d\theta
^{2}+\sin ^{2}\theta d\phi ^{2}\right) ,  \label{5}
\end{equation}%
where 
\begin{equation}
\alpha =\frac{3}{2\pi G\rho _{c}}.  \label{6}
\end{equation}%
is a constant, and $M$ is the ADM mass of the black hole. It can be shown
that the two metrics, (\ref{1}) and (\ref{5}), smoothly match at the
hypersurface $r=\tilde{r}_{0}a$, where both the first and second fundamental
forms are continuous across the hypersurface, provided that Eq. (\ref{6})
holds. Let us add that an observer in the exterior region measures an
energy-momentum tensor given by%
\begin{equation}
T_{\mu }^{\nu }=\frac{3\alpha M^{2}}{r^{6}}\left[ -1,-1,2,2\right] ,
\label{7}
\end{equation}%
which satisfies all energy conditions for $\alpha >0$, except the dominant
energy condition (DEC). To demonstrate this, we recall the definitions of
the energy conditions: i) the null energy condition (NEC) requires $\rho
+p_{i}\geq 0$, ii) the weak energy condition (WEC) requires $\rho \geq 0,$
and $\rho +p_{i}\geq 0$, iii) the strong energy condition (SEC) requires $%
\rho +p_{i}\geq 0,$ and $\rho +\sum p_{i}\geq 0$, and iv) the dominant
energy condition (DEC) requires $\rho \geq \left\vert p_{i}\right\vert $. By
defining the energy density and pressure components as $\rho =-T_{t}^{t}=%
\frac{3\alpha M^{2}}{r^{6}}$, $p_{r}=T_{r}^{r}=-\rho ,$ and $p_{\theta
}=p_{\phi }=T_{\theta }^{\theta }=T_{\phi }^{\phi }=2\rho $, it is
straightforward to verify that NEC, WEC, and SEC are satisfied, while DEC is
not.

The black hole referenced in (\ref{5}) is widely recognized in the
literature as the quantum-mechanically corrected black hole or quantum
Oppenheimer-Snyder (qOS) black hole. In \cite{R0} Cao et al. examined the
stability of the inner horizon of the qOS black hole. Using both a test
scalar field analysis and the generalized Dray-'t Hooft-Redmond relation,
the authors demonstrate that the inner (Cauchy) horizon of this black hole
is unstable, with flux and energy density diverging as free-falling
observers approach it. This instability leads to mass inflation and the
emergence of a null singularity, supporting the strong cosmic censorship
hypothesis. In \cite{R1}, Yang et al. investigate the effects of quantum
corrections on black hole shadows and stability within the framework of loop
quantum gravity. The study reveals that quantum corrections reduce the
radius of black hole shadows and analyzes the stability of these
quantum-corrected black holes by calculating quasinormal modes (QNMs). The
results indicate that the quantum-corrected black holes are stable against
scalar and vector perturbations, with the QNMs showing increased oscillation
frequencies and decreased damping rates compared to classical Schwarzschild
black holes. Recently, Dong et al. in \cite{2} conducted a detailed
investigation of this black hole, which introduces quantum corrections to
overcome the limitations of the classical model, especially the issue of the
Big Bang singularity. Their research delves into multiple facets of the
modified black hole metric, such as its thermodynamic properties, Hawking
radiation, quasi-normal modes, and the topological characteristics of photon
spheres and thermodynamic potentials. In another recent study Gong et al. in 
\cite{R2} investigated the QNMs spectra of the qOS black hole using scalar
perturbations. They found that the fundamental QNM modes of the qOS BH
exhibit two key properties: (1) a nonmonotonic behavior concerning the
quantum correction parameter for zero multipole number and (2) slower decay
modes due to quantum gravity effects. Moreover, in \cite{3} a novel
cosmological model called the quantum Oppenheimer-Snyder-Swiss Cheese
(qOSSC) model, which combines the qOS and quantum Swiss Cheese (qSC) models
within the framework of loop quantum cosmology, was presented. According to 
\cite{1}, the qOS model describes a collapsing matter ball inside a deformed
Schwarzschild black hole, while the qSC model represents a deformed
Schwarzschild black hole surrounded by a quantum-modified
Friedmann-Robertson-Walker (FRW) universe. Both models use the APS metric,
a semiclassical solution in loop quantum cosmology \cite{APS}.

In this article, we first focus on the qOS black hole in (\ref{5}).
Since this solution emerges from Einstein's equations with an accompanying
energy-momentum tensor, we suggest that nonlinear electrodynamics (NED)
could be the origin of the correction term. As we will demonstrate, the NED
model that aligns with our objectives is the power Maxwell law (PML),
introduced by Hassaine and Mart\'{\i}nez in \cite{HM1,HM2}. This model has
garnered attention across various fields of physics. For instance, black
holes in Lovelock theory coupled with MPL were explored in \cite{Lov}, and
wormholes in Einstein-Gauss-Bonnet gravity with PML were examined in \cite%
{Hendi}. Additionally, holographic superconductors within the framework of
PML were investigated in \cite{JL,Lee}, and numerous studies have explored
different aspects of the model \cite{M1,M2,M3,M4,M5,M6,M7,M8,M9,M10} (and
references therein).

After deriving the charged quantum-mechanically corrected black hole in the
first part of the paper, we proceed to reconstruct the qOS model to include
electric or magnetic charge. This new framework, referred to as the \textit{%
charged qOS model (cqOS)}, involves matching the exterior and interior
spacetimes at a massive and charged timelike spherical thin-shell.

\section{Quantum-mechanically corrected black hole in NED}

We now consider the action in Einstein's theory of gravity minimally coupled
to the PML NED, as proposed by Hassaine and Mart\'{\i}nez in \cite{2}, which
is expressed as%
\begin{equation}
I=\int d^{4}\sqrt{-g}\left[ \frac{R}{2\kappa }+\beta \mathcal{F}^{s}\right] ,
\label{8}
\end{equation}%
where $\kappa =8\pi G$, $\mathcal{F}=F_{\alpha \beta }F^{\alpha \beta }$ is
the Maxwell invariant, $\beta $ is a coupling constant, and $s\in 
\mathbb{R}
.$ The corresponding Einstein field equations are given by%
\begin{equation}
G_{\mu }^{\nu }=\kappa T_{\mu }^{\nu },  \label{9}
\end{equation}%
where the energy-momentum tensor $T_{\mu }^{\nu }$ takes the form%
\begin{equation}
T_{\mu }^{\nu }=4\beta \left( \frac{1}{4}\mathcal{F}^{s}\delta _{\mu }^{\nu
}-s\mathcal{F}^{s-1}F_{\mu \lambda }F^{\nu \lambda }\right) .  \label{10}
\end{equation}%
Additionally, with the electromagnetic two-form defined by%
\begin{equation}
\mathbf{F}=\frac{1}{2}F_{\mu \nu }dx^{\mu }\wedge dx^{\nu }  \label{11}
\end{equation}%
the nonlinear Maxwell equation becomes%
\begin{equation}
\frac{1}{\sqrt{-g}}\partial _{\mu }\left( \sqrt{-g}\frac{\partial }{\partial 
\mathcal{F}}\left( \beta \mathcal{F}^{s}\right) F^{\mu \nu }\right) =0.
\label{12}
\end{equation}%
Our objective is to determine the value of $s$ in equation (\ref{10}) such
that the energy-momentum tensor aligns with equation (\ref{7}). To achieve
this, we begin by considering a pure electric field described by 
\begin{equation}
\mathbf{F}=E\left( r\right) dt\wedge dr,  \label{13}
\end{equation}%
in conjunction with the line element 
\begin{equation}
ds^{2}=-f\left( r\right) dt^{2}+\frac{dr^{2}}{f\left( r\right) }+r^{2}\left(
d\theta ^{2}+\sin ^{2}\theta d\phi ^{2}\right) .  \label{14}
\end{equation}%
From this, we derive%
\begin{equation}
\mathcal{F}=-2E^{2}.  \label{15}
\end{equation}%
To ensure the action is physically meaningful, we redefine $\beta =\left(
-1\right) ^{s}\xi ,$ so that $\beta \mathcal{F}^{s}$ becomes $\xi \left( -%
\mathcal{F}\right) ^{s},$ where $\xi \in 
\mathbb{R}
.$ Furthermore, Maxwell's equation (\ref{12}) leads to%
\begin{equation}
E=\frac{C}{r^{\frac{2}{^{2s-1}}}},  \label{16}
\end{equation}%
where $C$ is an integration constant. Substituting Eq. (\ref{16}) into Eq. (%
\ref{10}), we obtain%
\begin{equation}
T_{\mu }^{\nu }=\left( 2\right) ^{s}\xi \frac{C^{2s}}{r^{\frac{4s}{^{2s-1}}}}%
diag\left( 1-2s,1-2s,1,1\right) .  \label{17}
\end{equation}%
By setting $\frac{4s}{^{2s-1}}=6,$ we find $s=\frac{3}{4},$ and thus%
\begin{equation}
T_{\mu }^{\nu }=\xi \frac{C^{3/2}}{\sqrt[4]{2}r^{6}}diag\left(
-1,-1,2,2\right) .  \label{18}
\end{equation}%
By comparing equations (\ref{18}) and (\ref{7}), we deduce that 
\begin{equation}
3\alpha M^{2}\equiv \xi \frac{C^{3/2}}{\sqrt[4]{2}}.  \label{19}
\end{equation}%
Consequently, the metric function for the electrically charged black hole
solution is derived as%
\begin{equation}
f\left( r\right) =1-\frac{2M}{r}+\frac{\xi C^{3/2}}{3\sqrt[4]{2}r^{4}}.
\label{20}
\end{equation}%
In (\ref{20}), $M$ represents the ADM mass of the black hole as measured by
an observer at $r\rightarrow \infty .$ Additionally, the integration
constant $C$ is related to the electric charge of the black hole, which can
be determined using Gauss's law%
\begin{equation}
q=\frac{1}{4\pi }\int_{S^{2}}\frac{\partial }{\partial \mathcal{F}}\left(
\xi \left( -\mathcal{F}\right) ^{s}\right) F^{tr}r^{2}\sin \theta d\theta
d\phi =s\xi 2^{s-1}C^{2s-1},  \label{R1}
\end{equation}%
where $s=\frac{3}{4}.$ Hence, we find $C=\frac{16\sqrt{2}}{9}\frac{Q^{2}}{%
\xi ^{2}},$ leading to the metric function%
\begin{equation}
f\left( r\right) =1-\frac{2M}{r}+\frac{3\alpha Q^{2}}{r^{4}},  \label{R2}
\end{equation}%
where $\alpha =\frac{2^{13/2}}{3^{5}}\xi ^{2}.$ In this reformulated metric
function, the electric charge $Q$ directly assumes the role of the mass $M$
in (\ref{8}), while the coupling constant $\xi $ mimics the quantum
correction parameter $\alpha .$

After determining $s$ in equation (\ref{8}) for a purely electric NED, we
now seek $s$ for a purely magnetic electromagnetic field. To this end, we
assume%
\begin{equation}
\mathbf{F}=P\sin \theta d\theta \wedge d\phi  \label{21}
\end{equation}%
which ensures that both the nonlinear Maxwell equation and the Bianchi
identity are trivially satisfied. Here, the constant parameter $P$
represents the magnetic charge, and the radial magnetic field is given by $B=%
\frac{P}{r^{2}}.$ Similar to the purely electric case, we redefine $\beta
=-\xi ,$ with $\xi \in 
\mathbb{R}
.$ The Maxwell invariant is then calculated as%
\begin{equation}
\mathcal{F}=\frac{2P^{2}}{r^{4}}  \label{22}
\end{equation}%
and the corresponding energy-momentum tensor becomes%
\begin{equation}
T_{\mu }^{\nu }=\xi \frac{2^{s}P^{2s}}{r^{4s}}diag\left(
-1,-1,-1+2s,-1+2s\right) .  \label{23}
\end{equation}%
By setting $4s=6,$ we obtain $s=\frac{3}{2},$ and thus equation (\ref{23})
simplifies to%
\begin{equation}
T_{\mu }^{\nu }=\frac{2^{3/2}\xi P^{3}}{r^{6}}diag\left( -1,-1,2,2\right) .
\label{24}
\end{equation}%
Comparing equations (\ref{24}) and (\ref{7}) yields%
\begin{equation}
2^{3/2}\xi P^{3}=3\alpha M^{2},  \label{25}
\end{equation}%
and the metric function for the magnetic black hole solution is found to be%
\begin{equation}
f\left( r\right) =1-\frac{2M}{r}+\frac{3\alpha P^{2}}{r^{4}}.  \label{26}
\end{equation}%
Here, in analogy with the electric solution, we introduce $\alpha =\frac{%
2^{3/2}\xi P}{9}.$ Furthermore, the magnetic charge corresponds to the mass
through the relation $P^{2}=M^{2}.$

To conclude this section, we note that both the electric and magnetic metric
functions, given respectively in Eqs. (\ref{R2}) and (\ref{26}), may admit
no horizon, one double horizon, or two horizons, depending on the
parameters' values. For the electric solution, the double horizon occurs
when $M=M_{c}=\frac{2}{\sqrt{3}}\left( \alpha Q^{2}\right) ^{1/4},$ with the
double horizon located at $r_{+}=\sqrt{3}\left( \alpha Q^{2}\right) ^{1/4}.$
For the magnetic solution, $M=M_{c}=\frac{2}{\sqrt{3}}\left( \alpha
P^{3}\right) ^{1/4},$ and the double horizon is located at $r_{+}=\sqrt{3}%
\left( \alpha P^{2}\right) ^{1/4}$. The subscript $c$ in $M_{c}$ denotes the
critical mass, such that for $M<M_{c},$ no horizon forms, and the spacetime
exhibits a naked singularity. On the other hand, for $M>M_{c},$ two horizons
exist: the event horizon and the Cauchy horizon. In \cite{2}, the authors
explored various aspects of the quantum-corrected black hole described by
Eq. (\ref{5}). These findings are equally applicable to the electric and
magnetic black hole solutions presented in this study.

\section{cqOS model}

In this section, we join the interior APS metric \cite{APS} to the exterior
charged quantum-mechanically corrected black hole model. The interior line
element is given by%
\begin{equation}
ds_{APS}^{2}=-d\tau ^{2}+a\left( \tau \right) ^{2}\left( d\tilde{r}^{2}+%
\tilde{r}^{2}d\Omega ^{2}\right) ,  \label{27}
\end{equation}%
while the exterior is described by Eq. (\ref{14}), where the metric function 
$f\left( r\right) $ is defined by either (\ref{R2}) or (\ref{26}). The two
metrics are glued at a spherical timelike hypersurface defined by $R\left(
\tau \right) =r=a\left( \tau \right) \tilde{r}_{0}$. The induced line
elements on the inner and outer sides of the hypersurface are respectively%
\begin{equation}
ds_{-}^{2}=-d\tau ^{2}+a\left( \tau \right) ^{2}\tilde{r}_{0}^{2}d\Omega
_{-}^{2},  \label{28}
\end{equation}%
and%
\begin{equation}
ds_{+}^{2}=\left( -f\left( r\right) \dot{t}^{2}+\frac{\dot{R}\left( \tau
\right) ^{2}}{f\left( r\right) }\right) d\tau ^{2}+R\left( \tau \right)
^{2}d\Omega _{+}^{2}.  \label{29}
\end{equation}%
Here, a dot denotes the derivative with respect to the proper time $\tau ,$
and the subscripts $\pm $ denote the inner ($-$) and outer ($+$) sides of
the hypersurface. By imposing $-f\left( r\right) \dot{t}^{2}+\frac{\dot{R}%
\left( \tau \right) ^{2}}{f\left( r\right) }=-1$ and $d\Omega
_{-}^{2}=d\Omega _{+}^{2},$ we obtain 
\begin{equation}
ds_{-}^{2}=ds_{+}^{2}=-d\tau ^{2}+a\left( \tau \right) ^{2}\tilde{r}%
_{0}^{2}d\Omega ^{2},  \label{30}
\end{equation}%
which satisfies the first Israel junction condition \cite{Isr1,Isr2}. The
second Israel junction condition (with $G=1$) requires%
\begin{equation}
\left[ K_{i}^{j}\right] -\left[ K\right] \delta _{i}^{j}=-8\pi S_{i}^{j},
\label{31}
\end{equation}%
where $\left[ K_{i}^{j}\right] =K_{i+}^{j}-K_{i-}^{j},$ with $K_{i+}^{j}$
and $K_{i-}^{j}$ representing the mixed extrinsic curvatures of the outer
and inner sides of the hypersurface, respectively. Similarly, $\left[ K%
\right] =$ $\left[ K_{i}^{i}\right] $ is the trace of $\left[ K_{i}^{j}%
\right] .$ Additionally, $S_{i}^{j}=diag\left[ -\sigma ,p,p\right] $ is the
surface energy-momentum tensor. In \cite{1}, where the exterior metric is
given by (\ref{5}), $\sigma =p=0,$ and the black hole mass is determined by
the interior energy-momentum tensor in (\ref{K1}), provided $\alpha $
satisfies Eq. (\ref{6}). Our detailed calculations yield (see \cite{Habib}
and the references therein)%
\begin{equation}
\sigma =\frac{1}{4\pi R}\left( 1-\sqrt{f\left( R\right) +\dot{R}\left( \tau
\right) ^{2}}\right) ,  \label{32}
\end{equation}%
and%
\begin{equation}
p=\frac{1}{8\pi }\left( \frac{1}{R}\left( 1-\sqrt{f\left( R\right) +\dot{R}%
\left( \tau \right) ^{2}}\right) -\frac{f^{\prime }\left( R\right) +2\ddot{R}%
}{2\sqrt{f\left( R\right) +\dot{R}\left( \tau \right) ^{2}}}\right) .
\label{33}
\end{equation}%
Next, we set $M=M_{ADS}+E$ and $Q^{2}=M_{ADS}^{2}+q^{2}$ for the
electrically charged black hole exterior spacetime, where 
\begin{equation}
M_{ADS}=\frac{4}{3}\pi \tilde{r}_{0}^{3}a^{3}\rho ,  \label{N}
\end{equation}%
is the total constant mass of the dust ball with radius $a\left( \tau
\right) \tilde{r}_{0}.$ Here $E$ is the asymptotic energy of the thin shell,
and $q$ is part of its electric charge (see \cite{Hod1,Hod2}). Knowing that
the dust energy density $\rho $ satisfies (\ref{3}) and $\alpha $ is given
by Eq. (\ref{6}), (\ref{32}) and (\ref{33}) become%
\begin{equation}
\sigma =\frac{1}{4\pi R}\left( 1-\sqrt{1-\frac{2E}{R}+\frac{3\alpha q^{2}}{%
R^{4}}}\right) ,  \label{34}
\end{equation}%
and%
\begin{equation}
p=\frac{1}{8\pi }\left( \frac{1}{R}\left( 1-\sqrt{1-\frac{2E}{R}+\frac{%
3\alpha q^{2}}{R^{4}}}\right) -\frac{\frac{2E}{R^{2}}-\frac{12\alpha q^{2}}{%
R^{5}}}{2\sqrt{1-\frac{2E}{R}+\frac{3\alpha q^{2}}{R^{4}}}}\right) .
\label{35}
\end{equation}%
In this setup, the mass of the thin shell is given by 
\begin{equation}
m=4\pi R^{2}\sigma =R\left( 1-\sqrt{1-\frac{2E}{R}+\frac{3\alpha q^{2}}{R^{4}%
}}\right) .  \label{36}
\end{equation}%
This equation implies%
\begin{equation}
E=m-\frac{m^{2}}{2R}+\frac{3\alpha q^{2}}{2R^{3}},  \label{37}
\end{equation}%
and after substituting $q^{2}=Q^{2}-M_{ADS}^{2},$ it becomes%
\begin{equation*}
E=m-\frac{m^{2}}{2R}-\frac{3\alpha M_{ADS}^{2}}{2R^{3}}+\frac{3\alpha Q^{2}}{%
2R^{3}}.
\end{equation*}%
Here, the asymptotic energy consists of four terms: the actual mass, the
gravitational energy of the shell, the gravitational energy of the dust
ball, and the electromagnetic energy. By solving $\frac{dE}{dR}=0,$ we find
the critical radius of the shell%
\begin{equation}
R_{eq}=\frac{3\sqrt{\alpha }\sqrt{Q^{2}-M_{ADS}^{2}}}{m}.  \label{38}
\end{equation}%
At this critical radius, we find $\frac{d^{2}E}{dR^{2}}=\frac{m^{5}}{%
27\alpha ^{3/2}q^{3}}>0,$ indicating that the asymptotic energy is
minimized, and the thin shell is stable. Consequently, the shell's radius
oscillates around this stable equilibrium radius without collapsing or
evaporating. The stable equilibrium radius of the timelike spherical
hypersurface (\ref{38}) depends on the mass $m$ and charge $Q$ of the shell,
as well as the constant mass of the dust ball $M_{ADS}$, and its existence
relies on the condition $Q>M_{ADS}.$ We note that this cqOS model reduces to
its charge-neutral version, qOS, introduced in \cite{1}. Finally, for the
magnetically charged case, one can substitute $Q$ by $P,$ and all results
remain valid.

\section{Conclusion}

In this study, we proposed a potential origin for the quantum correction of
the Schwarzschild black hole. This quantum-corrected black hole, previously
utilized by Lewandowski in \cite{1} to construct the regular qOS black hole
model, is characterized by the black hole mass $M$ and a correction
parameter $\alpha $. When $\alpha \rightarrow 0$, the model reverts to the
standard Schwarzschild black hole. By employing the NED framework developed
by Hassaine and Mart\'{\i}nez, known as the PML, we introduced two possible
sources for the correction term in the form of electric or magnetic charges.
These correspond to the PML NED model with $s=\frac{3}{4}$ and $\frac{3}{2}$%
, respectively. Additionally, we joined this charged black hole to an
interior spacetime represented by a dust ball governed by the APS metric.
The two spacetimes are joined at a timelike spherical hypersurface with mass 
$m$ and electric charge $Q$ or magnetic charge $P$. By minimizing the
asymptotic energy of the spacetime, we derived a stable equilibrium radius $%
R_{eq}$, around which the thin shell oscillates. This ensures the stability
of the entire spacetime, preventing collapse. The existence of this stable
equilibrium radius requires $Q>M_{ADS}$ for the electrically charged case
and $P>M_{ADS}$ for the magnetically charged thin shell.

\end{document}